\documentclass{elsart}
\usepackage{amsmath}
\usepackage{amssymb}
\usepackage{graphicx}
\journal{Physics Letters B}
\begin{document}
\runauthor{P. Arumugam, A. Ganga Deb and S.K. Patra}
\begin{frontmatter}
\title{Applicability of shape parameterizations for giant
dipole resonance in warm and rapidly rotating nuclei}

\author[iopb]{P. Arumugam\thanksref{aru}},
\author[sone]{A. Ganga Deb} and
\author[iopb]{S. K. Patra}

\address[iopb]{Institute of Physics, Sachivalaya Marg, Bhubaneswar
- 751 005, India.}

\address[sone]{Sonepur College, Subarnapur - 767017, Orissa, India.}

\thanks[aru]{\textit{E-mail address:} aru@iopb.res.in}

\date{August 16 2004}

\begin{abstract}
We investigate how well the shape parameterizations are applicable
for studying the giant dipole resonance (GDR) in nuclei, in the
low temperature and/or high spin regime.  The shape fluctuations
due to thermal effects in the GDR observables are calculated using
the actual free energies evaluated at fixed spin and temperature.
The results obtained are compared with Landau theory calculations
done by parameterizing the free energy. We exemplify that the
Landau theory could be inadequate where shell effects are
dominating. This discrepancy at low temperatures and high spins
are well reflected in GDR observables and hence insists on exact
calculations in such cases.
\end{abstract}

\begin{keyword}
Giant dipole resonance, thermal fluctuations, Landau theory, hot
rotating nuclei
\end{keyword}

\end{frontmatter}

In recent years considerable interest has been shown
\cite{Heck03,Came03,Brac03,Kusn03} to study the structural
transitions as a function of both angular momentum and temperature
in highly excited nuclei.  The Giant Dipole Resonance (GDR)
studies have been proved to be a powerful tool to study such hot
and rotating nuclei \cite{GDRREV} and recently the domain of GDR
spreads rapidly over different areas of theoretical and
experimental interest \cite{Maj95,Came01,Tson00,Dios01}.  The GDR
observations provide us information about the geometry as well as
the dynamics of nuclei even at extreme limits of temperature
($T$), spin ($I$) and isospin ($\tau$). In the past most of the
GDR measurements in hot nuclei were made at moderate and high $T$
\cite{GDRREV,Maj95,Came01,Tson00,Dios01,Gund90,Baum98,Kmie00}. Few
experiments have been carried out recently to study the GDR states
at low temperatures \cite{Heck03,Came03,Brac03}. Hence the
theories which were successful in the high $T$ regime should be
now scrutinized with the low temperature observations as well as
with theories incorporating properly the microscopic effects (such
as shell effects) which are dominant at low temperatures.  In a
macroscopic approach, while dealing with the thermal shape
fluctuations, free energy parameterizations such as Landau theory
\cite{ALHA,ALHAE,SELVP} are usually employed to do timesaving
calculations.  In this work we survey the applicability of Landau
theory by demanding consistency with exact calculations done
without any parameter fitting.

The theoretical approach we follow is of three fold with models
for 1) shape calculations, 2) relating the shapes to GDR
observables and 3) considering the shape fluctuations due to
thermal effects. For shape calculations we follow the
Nilsson-Strutinsky (NS) method extended to high spin and
temperature \cite{aruprc1}. The total free energy
($F_\mathrm{TOT}$) at fixed deformation is calculated using the
expression
\begin{equation}\label{FTOT}
F_\mathrm{TOT}=E_\mathrm{RLDM}+\sum_{p,n}\delta F \;.
\end{equation}
Expanding the rotating liquid-drop energy $E_\mathrm{RLDM}$ and
writing shell corrections in rotating frame \cite{NEER} leads to
\begin{equation}\label{FTOT1}
F_\mathrm{TOT}=E_\mathrm{LDM}+\sum_{p,n}\delta F^\omega +\frac12
\omega (I_\mathrm{TOT}+\sum_{p,n}\delta I)\;.
\end{equation}
The angular velocity $\omega$ is tuned to obtain the desired spin
given by
\begin{equation}
    I_\mathrm{TOT}=\Im_{rig}\omega +\delta I \;.
\end{equation}
The liquid-drop energy ($E_\mathrm{LDM}$) is calculated by summing
up the Coulomb and surface energies \cite{RAMS} corresponding to a
triaxially deformed shape defined by the deformation parameters
$\beta$ and $\gamma$.  The rigid-body moment of inertia
($\Im_{rig}$) is calculated with surface diffuseness correction
\cite{RAMS}. The shell correction ($\delta F^\omega$) is the
difference between the deformation energies evaluated with a
discrete single-particle spectrum and by smoothing (averaging)
that spectrum and is given by the relation
\begin{equation}
\label{dFT}\delta
F^\omega=F^\omega-\widetilde{F}^\omega=\left(\sum_{i=1}^\infty
e_i^\omega n_i-T\sum_{i=1}^\infty s_i\ \right)-\left(\sum_i
e_i^\omega\widetilde n_i-T\sum_i\widetilde s_i\right) \;,
\end{equation}
where the discrete quantities
\begin{equation}
\label{NI}n_i=\frac 1{1+\exp \left( \frac{e_i^\omega-\lambda
}T\right) }\;,
\end{equation}
$s_i=-\left[n_i\ln n_i-(1-n_i)\ln (1-n_i)\right]$, the Strutinsky
smeared quantities $\widetilde n_i=\int_{-\infty }^\infty
\widetilde f(x)\;n_i(x)\;d x$, $\widetilde s_i=\int_{-\infty
}^\infty \widetilde f(x)\;s_i(x)\;d x$ and $\widetilde f(x)$ is
the smearing function \cite{aruprc1}. Similarly the shell
correction corresponding to the spin is given by
\begin{equation}
\delta I=I-\widetilde{I}=\sum_{i=1}^\infty m_i n_i -
\sum_{i=1}^\infty m_i \widetilde{n}_i \;.
\end{equation}
The single-particle energies ($e_i^\omega$) and spin projections
($m_i$) are obtained by diagonalizing the triaxial Nilsson
Hamiltonian in cylindrical representation upto first twelve major
shells.

In a macroscopic approach, the GDR observables are related to the
nuclear shapes.  This is realized using a model \cite{THIAG}
comprising an anisotropic harmonic oscillator potential with
separable dipole-dipole interaction.  In this formalism the GDR
frequencies in laboratory frame are obtained as
\begin{equation}
\widetilde{\omega}_z=(1+\eta )^{1/2}\omega _z \;,
\end{equation}
\begin{eqnarray}
\nonumber
\widetilde{\omega}_2\mp \Omega&=&\left\{(1+\eta )\frac{\omega _y^2+\omega _x^2}%
2\;+\Omega^2 + \frac12\left[ (1+\eta )^2(\omega _y^2-\omega
_x^2)^2 \right. \right. \\ && \left. \left. +8\Omega ^2 (1+\eta
)(\omega _y^2+\omega _x^2)\right] ^{\frac 12}\right\} ^{\frac
12}\mp \Omega \;,
\end{eqnarray}
\begin{eqnarray}
\nonumber
\widetilde{\omega}_3\mp\Omega&=&\left\{ (1+\eta )\frac{\omega _y^2+\omega _x^2}%
2\;+\Omega ^2 -\frac 12\left[ (1+\eta )^2(\omega _y^2-\omega
_x^2)^2 \right. \right. \\&& \left. \left. +8\Omega ^2(1+\eta
)(\omega _y^2+\omega _x^2)\right] ^{\frac 12}\right\} ^{\frac
12}\mp\Omega \;,
\end{eqnarray}
where $\Omega$ is the cranking frequency, $\omega_x,\ \omega_y,\
\omega_z$ are the oscillator frequencies derived from the
deformation of the nucleus and $\eta$ is a parameter that
characterizes the isovector component of the neutron and proton
average field.  The GDR cross sections are constructed as a sum of
Lorentzians given by
\begin{equation}
\sigma (E_\gamma)=\sum_{i}
\frac{\sigma_{mi}}{1+\left(E_\gamma^2-E_{mi}^2\right)
^2/E_\gamma^2\Gamma_{i}^2} \;,
\end{equation}
where Lorentz parameters $E_m$, $\sigma _m$ and $\Gamma$ are the
resonance energy, peak cross-section and full width at half
maximum respectively. Here $i$ represents the number of components
of the GDR and is determined from the shape of the nucleus
\cite{THIAG,HILT}. The energy dependence of the GDR width can be
approximated by \cite{CARL}
\begin{equation}\label{PowLaw}
\Gamma_i \approx 0.026E_i ^{1.9} \;.
\end{equation}
The peak cross section $\sigma_m$ is given by
\begin{equation}
 \sigma_m=60\frac2\pi \frac {NZ}{A} \frac{1}{\Gamma} \;
 0.86(1+\alpha) \;.
\end{equation}
The parameter $\alpha$ which takes care of the sum rule is fixed
at 0.3 for all the nuclei considered in this work.  In most of the
cases we normalize the peak with the experimental data and hence
the choice of $\alpha$ has less effect on the results.  The other
parameter $\eta$ varies with nucleus so that the ground state GDR
centroid energy is reproduced. The choice for $^{84}$Zr is
$\eta=2.6$, for  $^{147}$Eu, $\eta=2.8$,  for $^{179}$Au,
$\eta=3.2$, and for $^{208}$Pb, $\eta=3.4$.  For calculating the
GDR width, only the power law (\ref{PowLaw}) is used in this work
and no ground state width is assumed.

When the nucleus is observed at finite excitation energy, the
effective GDR cross-sections carry information on the relative
time scales for shape rearrangements \cite{ALHAT} which lead to
shape fluctuations. In the case of hot and rotating nuclei, apart
from thermal shape fluctuations, there can be fluctuations in the
orientation of the nuclear symmetry axis with respect to the
rotation axis. The general expression for the expectation value of
an observable $\mathcal{O}$ incorporating both thermal and
orientation fluctuations is given by \cite{ALHA,ALHAO}
\begin{equation} \label{EqOFTF}
\langle \mathcal{O} \rangle_{\beta,\gamma,\Omega} = \frac{\int
\mathcal{D}[\alpha] \e^{-F(T,I;\beta,\gamma,\Omega)/T}
(\hat{\omega}\cdot\mathcal{I}\cdot\hat{\omega})^{-3/2}
\mathcal{O}} {\int
\mathcal{D}[\alpha]\e^{-F(T,I;\beta,\gamma,\Omega)/T}
(\hat{\omega}\cdot\mathcal{I}\cdot\hat{\omega})^{-3/2}} \;,
\end{equation}
where $\Omega=(\phi,\theta,\psi)$ are the Euler angles specifying
the intrinsic orientation of the system,
$\hat{\omega}\cdot\mathcal{I}\cdot\hat{\omega}=\Im_{x^\prime
x^\prime}\cos^2\phi\ \sin^2\theta + \Im_{y^\prime
y^\prime}\sin^2\phi\ \sin^2\theta + \Im_{z^\prime z^\prime}
\cos^2\theta$ is the moment of inertia about the rotation axis
$\hat{\omega}$ given in terms of the principal moments of inertia
$\Im_{x^\prime x^\prime},\ \Im_{y^\prime y^\prime},\ \Im_{z^\prime
z^\prime}$, and the volume element $\mathcal{D}[\alpha] = \beta^4
|\sin 3\gamma| \, d\beta \, d\gamma \, \sin\theta \, d\theta \,
d\phi$.

The study of thermal fluctuations by numerical evaluation of Eq.
(\ref{EqOFTF}) in general requires an exploration of five
dimensional space spanned by the deformation and orientation
degrees of freedom, in which a large number of points are required
in order to assure sufficient accuracy (especially at finite
angular momentum).  Hence certain parameterizations were developed
\cite{ALHAE,Orm97} to represent the free energy using functions
that mimic the behaviour of the NS calculation as closely as
possible. One such parametrization is the Landau theory of phase
transitions, developed by Alhassid and collaborators
\cite{ALHA,ALHAE,ALHAT,ALHAO}. Here the free energy is expanded in
terms of certain temperature dependent constants which are to be
extracted by fitting with the free energy calculations at fixed
temperatures from the NS method. Moreover, once the fits involving
free energy and moment of inertia are made for the non-rotating
case, the calculations can be extended to higher spins using the
relation \cite{ALHA}
\begin{equation} \label{EqFTI}
F(T,I;\beta,\gamma,\Omega)=F(T,\omega=0;\beta,\gamma)+\frac{(I+1/2)^2}
{2\ \hat{\omega}\cdot\mathcal{I}\cdot\hat{\omega}} \;.
\end{equation}
Hence this theory offers an economic parametrization to study the
hot rotating nuclei.  However the above expression carries the
shell corrections evaluated at $\omega=0$ all along to higher
spins.  This is not desirable as the single-particle levels
swiftly change positions with increasing spin, resulting in a
totally different shell structure.  We have employed Landau theory
in its extended form as given in Refs. \cite{ALHAE,aruprc1}.

Recently \cite{aruprc1,Ansa01} few calculations have been done by
performing the thermal fluctuation calculations exactly by
computing the integrations in Eq. (\ref{EqOFTF}) numerically with
the free energies and the observables being calculated ``exactly''
at the integration (mesh) points. In this way the calculations can
be done more accurately without using any parametrization and
consequent fitting.  In this work we have performed such
calculations, however, neglecting the orientation fluctuations.
This enables us to perform the integration in the deformation
space only which at present is two dimensional having the
deformation parameters $\beta$ and $\gamma$.  The role of
orientation fluctuations is negligible while calculating the
scalar observables \cite{aruprc1,Orm97a} such as the GDR cross
section and width. The derivation \cite{Orm97a} of Eq.\
\ref{EqOFTF} employs the assumption of Eq. (\ref{EqFTI}) and here
we discuss the consequences as the free energy now takes the form
of Eq.\ (\ref{FTOT}).  The partition function at fixed angular
momentum $I$ is obtained as \cite{Orm97a}
\begin{equation}\label{EqZI}
    Z_I(T)=\frac{2I+1}{2\pi iT^2}\int \mathcal{D}[\alpha]
    \left[ \int_{-i\infty }^{i\infty }d\omega\ \omega \ e^{-[\omega
(I+1/2)+F^{\omega}]/T}\right]\;.
\end{equation}
The exponent in the above integral can be taken as
$-F_\mathrm{TOT}/T$ in our case. Letting
$\hat{\omega}\cdot\mathcal{I}\cdot\hat{\omega}=\Im_{z^\prime
z^\prime}=\Im_{\mathrm{TOT}}=\Im_{rig}+\delta\Im$ and following
the definitions given in Ref. \cite{NEER}, we have
\[
d\left[e^{-F_{\mathrm{TOT}}/T}\right]=\Im_{\mathrm{TOT}}\ \omega \
e^{-F_{\mathrm{TOT}}/T}\ d\omega\;.
\]
In the limit $d\Im_{\mathrm{TOT}}/d\omega\rightarrow 0$, the
integration over $\omega$ in Eq. (\ref{EqZI}) is analytically
solvable leading to an expression having exactly same form as Eq.
(\ref{EqOFTF}) with
$\hat{\omega}\cdot\mathcal{I}\cdot\hat{\omega}$ replaced by
$\Im_{\mathrm{TOT}}$.  The shell corrections to moment of inertia
are significant for spherical shapes which are however suppressed
by the term $\beta^4$ in $\mathcal{D}[\alpha]$ and for $\beta \geq
0.1$ we have $\Im_{\mathrm{TOT}}\approx \Im_{rig}$. At high spins
also the limiting condition is very much valid as the well
deformed shapes are more favoured. It has already been shown
\cite{Orm97a} that this transformation from frequency to angular
momentum leads to a pre-factor to the volume element of the
integral.  However we show later that this factor has not much
role to play in the practical calculations.  While performing
fluctuation calculations in this way, the free energy at any given
spin is obtained by tuning the cranking frequency to get the
desired spin.  In such case the projection is not necessary and
has been neglected in some similar calculations \cite{Ansa01}.

\begin{figure*}
\centering
\includegraphics[width=0.8\textwidth, clip=true]{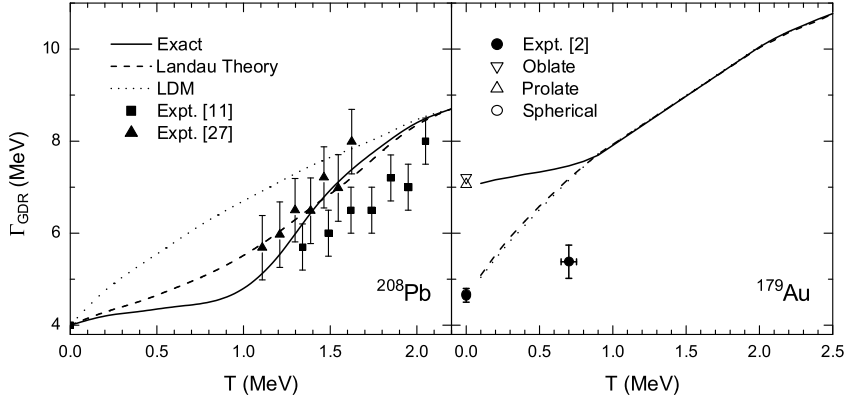}
\caption{GDR width for the nuclei $^{208}$Pb and $^{179}$Au. The
results obtained using liquid drop model (dotted line), Landau
theory (dashed line) and the exact calculations (solid line) are
compared.  \textit{Left:} Experimental data represented by solid
squares are taken from ref. \cite{Baum98} and the revised data
\cite{Kusn98} are represented by solid triangles.  \textit{Right:}
The different curves carry same meaning as in the left panel.  At
$T=0$ MeV, the widths calculated assuming oblate
($\beta_2=-0.22$), prolate ($\beta_2=0.24$) and spherical shapes
are denoted by the open symbols down-triangle, up-triangle and
circle respectively. The experimental value \cite{Came03} at $T=0$
MeV is the intrinsic (spherical) width $\Gamma_0$ and at $T=0.7$
MeV is the total width having $\Gamma_0=5$ MeV and $\beta_2=0.1$.
Both these values are represented by solid circles.}
\end{figure*}

Now we compare our calculated results obtained by using the
extended Landau theory and the exact method.  The calculations are
performed with 1) the liquid-drop model (LDM) free energies and
Landau theory, 2) NS free energies and Landau theory and 3) NS
free energies with exact treatment of fluctuations.

\begin{figure*}
\centering
\includegraphics[width=0.85\textwidth, clip=true]{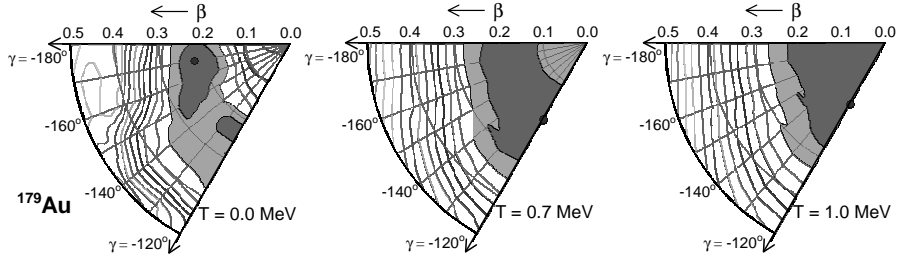}
\caption{Potential energy surfaces for $^{179}$Au at $T=0.0$, 0.7
and 1.0 MeV, calculated using the Nilsson-Strutinsky method.  The
contour line spacing in 0.5 MeV, the most probable shape is marked
by a solid circle and the first two minima are shaded.  The
barrier between the co-existing shapes diminish with increase in
temperature and the nucleus becomes $\gamma$-soft at $T\sim 0.7$
MeV.}
\end{figure*}

In Fig.\ 1 we present the calculated GDR widths of $^{208}$Pb and
$^{179}$Au at $\omega=0$ along with experimental results.  In the
case of $^{208}$Pb, strong shell corrections for spherical shape
results in large difference in the deformation energies between
spherical and deformed configurations.  This leads to attenuation
of thermal fluctuations at lower temperature and hence the
obtained widths are much lower when compared to liquid-drop model
results.  The magnitude of this attenuation comes out to be
different in methods 2 and 3.  More discussions and comparison
with other reported results can be found in Ref. \cite{aruprc1}.
Recently \cite{Came03} the thermal fluctuation calculations of
$^{208}$Pb were extrapolated to interpret low $T$ measurement of
$\Gamma_{GDR}$ in $^{179}$Au.  This extrapolation was assumed to
give the lower limit for the $\Gamma_{GDR}$.  Our calculations as
shown in Fig.\ 1 do not favour any such interpretation.  In fact
the situation in $^{179}$Au is totally different from that of
$^{208}$Pb as now deformed shapes are more favoured.  The ground
state deformation comes out in our calculations as $\beta_2=-0.22$
which is slightly larger than $\beta_2=-0.17$ obtained by
relativistic mean field theory (with NL3 parameter set) and
Hartree-Fock-Bogoliubov calculations \cite{HFB}. $\Gamma_{GDR}$
calculated for this deformation at $T=0$ MeV lie very close to
those obtained with exact thermal fluctuations at $T=0.1$ MeV.
Moreover coexisting oblate and prolate shapes also prevail as we
observe in the potential energy surfaces (PES) given in Fig.\ 2.
We infer from the PES that the deformed shapes only are favoured
even at $T=0.7$ MeV where we see $\gamma$-softness. As the
$\Gamma_{GDR}$ of deformed shapes are always higher than spherical
ones, the calculation without shell effects (LDM), which favour
spherical shapes, give the lower limit for $\Gamma_{GDR}$ in
deformed nuclei.  Hence the inclusion of shell corrections will
only increase $\Gamma_{GDR}$ and could not account for the
suppression of $\Gamma_{GDR}$ in $^{179}$Au.

A similar situation has been observed in $^{120}$Sn \cite{Heck03}
also where the thermal fluctuation model, even with shell effects
included, could not explain the suppression of $\Gamma_{GDR}$ at
low $T$. Our exact calculations (see Fig.\ 4 of Ref.
\cite{aruprc1}) also could not explain this anomaly.  It has been
speculated \cite{Thoen04} that this suppression could be a general
feature of all nuclei, independent of shell effects.  The possible
cause of this could be the pairing correlations which can be
significant at low $T$.  The inclusion of fluctuations in the
pairing field \cite{Dang03} or fully microscopic calculations
\cite{Stor04} considering the temperature dependence of spreading
width could explain the low width of $^{120}$Sn at low $T$.  The
same could be the case of $^{179}$Au also.  In the present work,
we wish to emphasis on the problems in using the shape
parameterizations apart from these effects.

Also we infer from the calculations of $^{208}$Pb and $^{179}$Au,
if the shell effects are strong leading to a crisp or multiple
minima, the Landau theory could not account for it.  This
discrepancy can be partially ascribed to the parameterization
itself and the fitting procedure involved.  It was already
suggested \cite{ALHAE} that at low temperatures the least square
fit for Landau constants could be erroneous and techniques like
uniform mapping should be adopted.  The new parameterization
\cite{Orm97} may solve some of these issues despite having some
inconsistencies at higher temperatures.  However in both the
methods involving shape parameterization, the high spin
calculations are done using Eq. (\ref{EqFTI}).  Due to the
complexity of shell structure at high spins, the difference in
shell effects with changing spin may be crucial especially at
lower temperatures.

\begin{figure*}
\centering
\includegraphics[width=0.8\textwidth, clip=true]{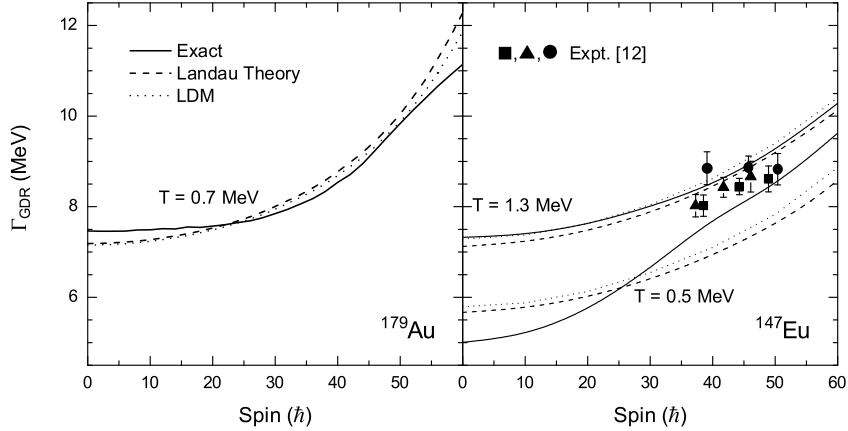}
\caption{Spin dependance of GDR width in $^{179}$Au and
$^{147}$Eu. The results obtained using liquid drop model
(dash-dotted line), Landau theory (dashed line) and the exact
calculations (solid line) are compared. 
\textit{Right:} The solid circle, solid square and solid triangle
correspond to experimental data \cite{Kmie00} at beam energies
170, 165 and 160 MeV respectively.  These energies correspond to
temperatures from 1.2 to 1.4 MeV \cite{Kmie00}.}
\end{figure*}

In our previous work \cite{aruprc1} we investigated the spin
dependence of $\Gamma_{GDR}$ in $^{120}$Sn at $T=1.8$ MeV (See
Fig.\ 5 of Ref.\ \cite{aruprc1}) and found the Landau theory to
perform well even at spins upto 70$\hbar$.  This is understandable
as there is no spin dependence of shell effects in this case and
the temperature is sufficiently high where the Landau theory works
well.  Our results for $^{179}$Au at $T=0.7$ MeV and at different
spins are shown in Fig.\ 3.  Here also the spin dependence of
shell effects are found to be insignificant.  We extended our
study to $^{147}$Eu for which experimental data is available for
$T\sim1.3$ MeV.  Our results are shown in Fig.\ 3 where we can see
that at $T=1.3$ MeV there is not much difference between the
results of Landau theory and exact calculations.  Also the widths
are very much similar to those obtained using LDM as the proton
and neutron shell corrections are weak and they act against
themselves. This trend continues even at spins up to 60$\hbar$.
However at $T=0.5$ MeV, the situation is drastically different as
the spin-driven shell effects play their role.  At $\omega=0$, the
shell correction is of the order of 2 MeV and hence the three
methods give different results.  The shell corrections decrease
with the increasing spin, at 40$\hbar$ the equilibrium shape
acquires deformation $(\beta,\gamma)=(0.3,-180^\circ)$ and at
60$\hbar$, it is $(0.3,-130^\circ)$.  The effect of these sharp
changes survive thermal fluctuations in the exact calculations and
is averaged out in the Landau theory as well as the LDM
calculations. It has to be noted that the Landau theory results
carry the shell corrections calculated at $\omega=0$ all along.
From the above arguments it is clear that one cannot substantiae
the success of Landau theory in the cases of $^{120}$Sn and
$^{179}$Au.  For certain nuclei like $^{147}$Eu the spin-driven
shell effects can be crucial at low $T$.

\begin{figure}
\centering
\includegraphics[width=0.4\textwidth, clip=true]{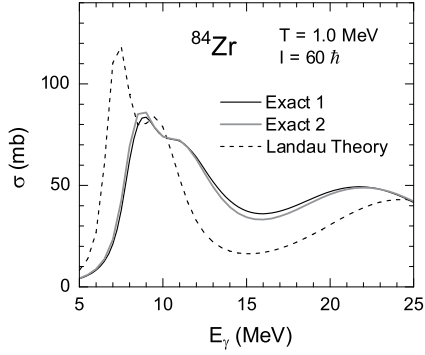}
\caption{GDR cross sections for the nuclei $^{84}$Zr at $T=1.0$
MeV and $I=60\hbar$ with shape fluctuations using Landau theory
(dashed line), exact method with the factor
$\Im_{\mathrm{TOT}}^{-3/2}$ (Exact 1, solid black line) and exact
method without that factor (Exact 2, solid grey line).}
\end{figure}

In Fig.\ 4 we show for the hot rotating $^{84}$Zr nucleus, the
results of our GDR cross section calculations.  It has been
observed that in neutron-deficient Zr isotopes, spin-driven shell
effects are stronger leading to a sharp shape transition at lower
temperatures \cite{aruprc1}.  It is evident from the Fig.\ 4 that,
in certain nuclei at higher spins the spin-driven shell effects
may play vital role even at $T\sim 1$ MeV.   Also we show in Fig.\
2 the impact of including the pre-factor
$\Im_{\mathrm{TOT}}^{-3/2}$ in the thermal fluctuation integral.
This is the only case presented in this work where we could notice
a slight effect of the factor.  One can observe that even at
extreme limits of spin the pre-factor has practically no effect as
the dominating role is played by the exponential term
$\exp(-F_{\mathrm{TOT}}/T)$ which has exact temperature and spin
dependence.

To summarize, in this work the thermal fluctuations are dealt in
an exact way without any parameter fitting.  Comparison of our
present approach with the thermal fluctuation model comprising
Landau theory suggests that the shape parameterizations could be
insufficient for GDR calculations in the presence of strong shell
effects. The discrepancies are shown in certain nuclei at two
situations namely at very low temperatures ($T < 1$ MeV) and at
moderate temperatures and high spin ($T\sim 1$ MeV, $I \gtrsim
30\hbar$). The latter case is ascribed to spin driven shell
effects, which the existing shape parametrizations do not account
for.  The present study necessitates exact treatment of
fluctuations in these cases where there is experimental and
theoretical focus recently.

\end{document}